\documentclass[aps,pra,10pt,superscriptaddress,showpacs,twocolumn,nofootinbib,floatfix,amsmath,amssymb]{revtex4-1}
\usepackage{mathtools}
\usepackage[usenames,dvipsnames]{xcolor}
\usepackage{color}
\usepackage{braket}
\usepackage{bm}
\usepackage{graphicx}
\usepackage{hyperref}
\usepackage[T1]{fontenc}

\begin{document}
\title{Few strongly interacting ultracold fermions in one-dimensional traps of different shapes}

\author{Daniel P{\k e}cak}
\email{pecak@ifpan.edu.pl}
\affiliation{Institute of Physics, Polish Academy of Sciences,  Aleja Lotnikow 32/46, PL-02668 Warsaw, Poland}

\author{Tomasz Sowi\'nski}
\affiliation{Institute of Physics, Polish Academy of Sciences,  Aleja Lotnikow 32/46, PL-02668 Warsaw, Poland}

\begin{abstract}
The ground-state properties of a few spin-1/2 fermions with different masses and interacting via short-range contact forces are studied within an exact diagonalization approach. It is shown that, depending on the shape of the external confinement, different scenarios of the spatial separation between components manifested by specific shapes of the density profiles can be obtained in the strong interaction limit. We find that the  ground-state of the system undergoes a specific transition between orderings when the confinement is changed adiabatically from a uniform box to a harmonic oscillator shape. We study the properties of this transition in the framework of the finite-size scaling method adopted to few-body systems. 
\end{abstract}

\maketitle

\section{Introduction}
With recent experiments on several particles confined in a one-dimensional optical trap (for fermions as well as for bosons), quantum engineering has entered a completely new, so far unexplored, area of strongly correlated quantum systems \cite{paredes2004tonks,kinoshita2004observation,Murmann2015AntiferroSpinChain,Kaufman2015Entangling}. In these extremely sophisticated experiments it is possible to control  the total number of particles, their mutual interactions, and the shape of external potential with very high accuracies \cite{Kaufman2015Entangling,haller2009realization,serwane2011deterministic,Murmann2015AntiferroSpinChain,Gaunt2013BECinBox}. As a consequence, a deep analysis of many properties of one-dimensional few-body systems is performed experimentally. For example, fermionization of distinguishable particles \cite{zurn2012fermionization}, pairing for attractive forces \cite{zurn2013Pairing}, ground-state properties in double-well schemes \cite{Murmann2015DoubleWell}, or the formation of the Fermi Sea \cite{Kohl2005,wenz2013fewToMany} have been observed already. In parallel, on a theoretical level many interesting results have been obtained under the assumption that particles are confined in a harmonic trap \cite{Sowinski2015Pairing,DAmico2015Pairing,Moszynski2015FewManyFermions,Moszynski2015ManyFermions,GarciaMarch2014ThreeBosons,GarciaMarch2014Localization,GarciaMarch2015Conductivity,Malet2015DensityFunctional,BjerlinReimann2016Higgs,Gharashi2015ImpurityUpperBranch,Zinner2015Anyons,YangCui2016SpinChain,Decamp2016SixFlavors,Drut2014Phase,Drut2015Phase,Valiente2014Bosons1D,Valiente2014Multi,Koscik2012,Koscik2016Bipartite}. They are awaiting experimental confirmations. Some results also for other confinements, like the double-well potential, have been discussed recently \cite{Dobrzyniecki2016DoubleWell,Sowinski2016Diffusion} and the dynamical properties of such systems have been analyzed.

Apart from a few exceptions \cite{loft2014variational,Blume2008SmallImbalanced,Mehta2015AttractiveImpurity,Mehta2014BornOppenheimer,Blume2010UniversalityBreakdown,Blume2012FewBodyTraps,daily2012thermodynamics,Dehkharghani2016Impenetrable}, it has commonly been assumed that particles of different kinds have the same mass and the main impact on properties of the system comes from an imbalance of the number of particles. However, recently it was shown that for particles confined in a harmonic trap, the mass difference between different fermionic components leads to their spatial separation if interactions are strong enough \cite{Pecak2016Separation}. The mechanism was shown to be universal with respect to the number of particles and also very robust to external perturbations. A remaining open question concerns the properties when different shapes of the trap are considered. This question is interesting also from an experimental point of view, since shape-manipulation is one of the standard experimental methods that are well controlled in laboratories. Recently, it was even possible to perform the first Bose-Einstein condensation in a purely uniform box confinement \cite{Gaunt2013BECinBox}. Motivated by this background, here we explore the properties of a spatial separation mechanism for a two-flavoured mixture of fermions confined in a one-dimensional trap with a tunable shape. We show that, depending on the shape, in the strong interaction limit spatial separation in the many-body ground-state may occur for either the lighter or the heavier component. Moreover, the system undergoes a kind of critical transition that is induced by an adiabatic change of the external potential. This mechanism appears to be very general and it is present always whenever fermions of different mass are being considered. We believe that our results may shed some light on the quantum magnetism \cite{Jo2009itinerant,LiuHuDrummond2010ThreeAttractive,Pilati2014Ferromagnetism,bugnion2013ferromagnetic,GarciaMarch2013SharpCrossover,cui2014ground,Massignan2015MagnetismMixtures,Dehkharghani2016Magnetism} and the role of mass imbalance in spatial separation of the density profiles \cite{JasonHo2013PhaseSeparation,Fratini2014ZeroTemp}.

The article is organized as follows. In an introductory Section \ref{Sec:TheSystem} we describe the system to be studied and we define the tunable shape of the external trap that will be considered in further analysis. Then, in Section \ref{Sec:ExactDiag} we briefly summarize the exact diagonalization method -- our main tool for studying different properties of few-body problems. The spectral properties of the few-body Hamiltonian from the point of view of different mass components as well as different trap shapes are studied in Section \ref{Sec:Spectral}. Subsequently, in Section \ref{Sec:Uniform} we focus on properties of the ground-state of the system and we discuss the spatial separation of density profiles induced by different masses in a uniform box potential. We also outline the similarities and differences in comparison to harmonic confinement. Section \ref{Sec:EqualMass} emphasizes the fundamental differences regarding single-particle densities between systems with the same and with different masses of the components. In this section, basing on numerical results, we also postulate that for any confinement one of two types of separation will always occur in the system when particles of different flavours have different masses. This observation leads us to make a numerical study of the transition between different density orderings in Section \ref{Sec:Transition}. In that Section we adopt the well known finite-size scaling method to a few-body system. Finally, we conclude in Section \ref{Sec:Conclude}.

\section{The system under study} \label{Sec:TheSystem}
In this paper we consider a two-flavour mixture of several ultra-cold fermionic atoms confined in an effectively one-dimensional external potential. Experimentally, a one-dimensional geometry is obtained by applying a very strong harmonic confinement in the two remaining spatial directions\cite{zurn2012fermionization,wenz2013fewToMany,serwane2011deterministic}. Depending on the experimental realization, atoms in the two flavours can have the same or different masses. The latter system is realized simply by trapping different chemical elements. The most promising fermionic mixture of this type is the lithium-potassium combination. Obtaining a mixture of fermions of the same mass is a more sophisticated procedure and can be achieved when two different nuclear spin projections of the same element are under control. A typical example is the mixture of two different $^6$Li atoms with total atomic spin belonging to the spin-$3/2$ and spin-$1/2$ representations, respectively. Regardless of the situation, in both scenarios particles of different flavours can be treated as fundamentally distinguishable, i.e. each fermion always belongs to one of the two flavours and during the whole experiment its nature cannot be changed \cite{serwane2011deterministic}. This is a kind of superselection principle originating in the observation that interactions between atoms cannot change neither the mass of the atoms nor the spin projection of their nuclei. 

\begin{figure}
 \resizebox{0.48\textwidth}{!}{%
 \includegraphics{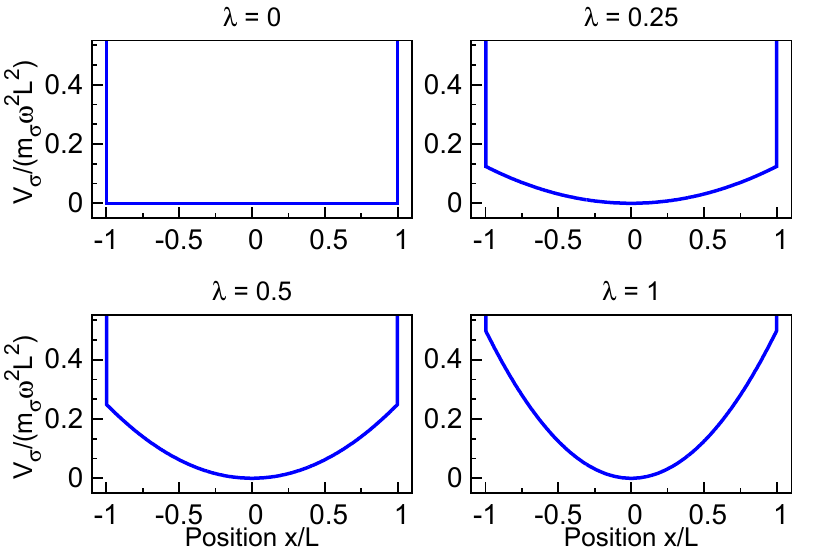}
 }
 \caption{The shape of the potentials $V_{\sigma}(x,\lambda)$ for different values of the parameter $\lambda$ in natural units of a given flavour. For $\lambda=0$ a uniform box potential is restored. For increasing $\lambda$, the confinement transforms to the standard harmonic oscillator.
  \label{fig:fig1} }
\end{figure}
It is a very good approximation to assume that ultra-cold fermions of different kinds interact only via spherically symmetric forces modeled by a zero-range $\delta$-like potential \cite{PethickSmith}. In this approximation, fermions belonging to the same flavour do not interact at all due to the antisymmetry of the wave function when written in terms of relative positions. In this approximation the Hamiltonian of the system reads 
\begin{equation} \label{eq:hamiltonian} 
\begin{split}
\hat{\cal H} = \sum_{i=1}^{N_{\downarrow}} \left[ - \frac{\hbar^2}{2 m_{\downarrow}} \frac{\partial^2}{\partial x_i^2} + V_{\downarrow}(x_i,\lambda) \right] + \\
 + \sum_{i=1}^{N_{\uparrow}} \left[ - \frac{\hbar^2}{2 m_{\uparrow}} \frac{\partial^2}{\partial y_i^2} + V_{\uparrow}(y_i,\lambda) \right] 
 + g_{\mathrm{1D}} \sum_{i,j=1}^{N_{\downarrow},N_{\uparrow}} \delta(x_i - y_j),
\end{split}
\end{equation}
where $V_{\sigma}(x,\lambda)$ is an external potential acting on fermions $\sigma$. We model the external potential as follows:
\begin{equation} \label{eq:extPotentialV}
V_{\sigma}(x,\lambda) =
\left\{
  \begin{array}{ll}
    \frac{1}{2} \lambda m_{\sigma} \omega^2  x^2   & \mbox{if } |x| < L \\
    \infty & \mbox{if } |x| > L,
  \end{array}
\right.
\end{equation}
where $\lambda$ is a dimensionless geometric parameter that determines the shape of the trap. For clearness, we use different letters for positions of particles belonging to different components. 

The confinement reproduces a uniform box potential of length $2 L$ in the limit of $\lambda \rightarrow 0$ and a cropped harmonic oscillator trap of frequency $\omega$ in the limit of $\lambda \rightarrow 1$. Of course, in the latter case hard walls affect and modify the single-particle eigenstates of the Hamiltonian. However, for low excited states and for a large enough $L$, the difference between an exact harmonic oscillator potential and one modeled by $V_{\sigma}(x,1)$ can be neglected. This conclusion comes from the observation that the wave functions of the harmonic oscillator decay exponentially and do not penetrate the regions in the vicinity of the hard walls of the uniform box \cite{Drut2015HardWall}.

In Fig.~\ref{fig:fig1} we schematically show the shape of the external potential for different values of $\lambda$ in natural units of a given flavour.
It is worth noticing that potential (\ref{eq:extPotentialV}) seems to be quite natural from an experimental point of view. It resembles the technique of turning off a harmonic oscillator potential in the presence of an additional uniform potential with hard walls \cite{Gaunt2013BECinBox}. Nevertheless, we have also checked a few other scenarios of crossover from an uniform box to a harmonic trap and found that the results described here do not depend qualitatively on these details. 

The effective interaction coupling strength $g_{\mathrm{1D}}$ is related to its three-dimensional counterpart and can be obtained by integrating out two remaining degrees of motion\cite{Olshanii1998}. From the point of view of our model the important information is that the interaction strength can be tuned experimentally over the whole range of its possible values, i.e. from minus to plus infinity \cite{Thalhammer2006, tiecke2010Feshbach6Li40K, Wille6Li40K}. Note that in contrast to higher dimensions, in a one-dimensional case, the Dirac $\delta$ function is a well defined self-adjoint Hermitian operator and can be used without any regularization\cite{Busch1998}. 

For a given shape of the confinement $\lambda$, we numerically find single-particle states $\phi^{(\lambda)}_{n\sigma}(x)$ and corresponding energies $E^{\mathrm{(\lambda)}}_{n\sigma}$ with a direct diagonalization of the single-particle Hamiltonian
\begin{equation} \label{eq:spHams}
{H}^{(\lambda)}_\sigma = - \frac{\hbar^2}{2 m_{\sigma}} \frac{\mathrm{d}^2}{\mathrm{d}x^2} + V_{\sigma}(x,\lambda) 
\end{equation}
The diagonalization is performed in the position domain on a dense grid with spacing $\delta x$. In this representation any single-particle Hamiltonian has a simple tridiagonal form. Therefore, a diagonalization is straightforward with standard numerical recipes \cite{NumRec1992}. It is quite obvious that along with decreasing $\delta x$, eigenstates and their eigenenergies converge to exact values. Here, to make numerical analysis possible, we assume that convergence is achieved when the relative numerical error of a number $n_{\mathrm{cutoff}}$ of the lowest states is smaller than $1\%$. The states $\phi^{(\lambda)}_{n\sigma}(x)$ serve as the basis for further many-body analysis.

In the limiting case of the harmonic oscillator ($\lambda\rightarrow1$), the single-particle eigenfunctions of both flavours are related by the following scaling: $\phi^{(1)}_{n\uparrow}(x) = (m_\uparrow/m_\downarrow)^{1/4}\,\phi^{(1)}_{n\downarrow}(\sqrt{m_\uparrow/m_\downarrow}\,x)$. This means that the wave functions of the heavier particles are more localized in the center of the trap. In this case, the eigenenergies do not depend on the mass of the particle and they depend linearly on the main quantum number $n=1,2,3,...$:
\begin{equation} \label{EnergyHO}
E^{\mathrm{(1)}}_{n} = \hbar \omega \left(n - \frac{1}{2}\right).
\end{equation}
Note, that for consistence of the whole analysis, we enumerate the single-particle states in such a way that the ground state is denoted by $n=1$ and not by $n=0$ as usually used in the literature for the harmonic oscillator problem.
It is also worth noticing that for a high enough excitation $n$, corrections from the hard-wall constraints become relevant. As explained before, to avoid this problem in our numerical approach, we set the size of the hard-wall box large enough to assure that the single-particle states that are appreciably occupied are not disturbed. We have checked that for our choice of $L$, the results of a pure harmonic oscillator confinement are restored for $\lambda=1$. Therefore, in the following we will treat $\lambda=1$ as a pure harmonic oscillator confinement.

In the opposing case of a uniform box potential ($\lambda=0$), the shapes of the wave functions do not depend on the mass and they have the well known form
\begin{equation}
\phi^{(0)}_{n\sigma}(x) =\sqrt{\frac{1}{L}}\sin\left[\frac{n \pi (x+L)}{2 L}\right].
\end{equation}
However, in this case, the single-particle eigenenergies depend on mass and the quantum number $n=1,2,3,...$
\begin{equation} \label{EnergyBox}
E^{\mathrm{(0)}}_{n\sigma} = \frac{\hbar^2 \pi^2 n^2}{8 m_{\sigma} L^2} \propto n^2.
\end{equation}

In what follows we will express all quantities in harmonic oscillator units of the spin-$\downarrow$ particles, i.e. all lengths are measured in units of $\sqrt{\hbar/(m_{\downarrow} \omega)}$, energies in $\hbar\omega$, momenta in units of $\sqrt{\hbar m_{\downarrow} \omega}$, etc. We also introduce the dimensionless mass ratio parameter $\mu=m_{\uparrow}/m_{\downarrow}$. This is substantially greater than unity for the lithium-potassium mixture, $\mu=40/6$.  In these units, the single-particle Hamiltonians \eqref{eq:spHams} have the form
\begin{equation} 
H^{(\lambda)}_\downarrow = -\frac{1}{2}\frac{\mathrm{d}^2}{\mathrm{d}x^2} + \frac{1}{2}\lambda x^2, \qquad H^{(\lambda)}_\uparrow = -\frac{1}{2\mu}\frac{\mathrm{d}^2}{\mathrm{d}x^2} + \frac{\mu}{2}\lambda x^2.
\end{equation}

To make the later analysis clear, we fix the size of the system in such a way that the single-particle spectra of the extreme Hamiltonians (i.e. those for a box trap and for a harmonic oscillator potential) have energy gaps of the same order of magnitude  i.e. $E^{\mathrm{(0)}}_{2\downarrow}-E^{\mathrm{(0)}}_{1\downarrow}\approx E^{\mathrm{(1)}}_{2}-E^{\mathrm{(1)}}_{1}$, which corresponds to the following condition
\begin{equation}\label{eq:orderOfMagnitude}
 1 \approx \frac{3 \hbar \pi^2}{8 m_{\downarrow}\omega L^2}.
\end{equation}
This condition determines an appropriate size of the system for numerics,  $2L\approx3.9\sqrt{\hbar/(m_{\downarrow}\omega)}$. To make sure that the walls do not noticeably affect the single-particle densities in the case of the cropped harmonic potential, we set the position of the walls to a larger value, $2L=7 \sqrt{\hbar/(m_{\downarrow}\omega)}$. With this condition, the energy gaps are still of the same order of magnitude.

For our numerical purposes it is convenient to rewrite the Hamiltonian (\ref{eq:hamiltonian}) in a dimensionless form in the second quantization formalism as follows:
\begin{equation} \label{eq:ham}
\begin{split}
 \hat{\cal H} = \sum_\sigma\int_{-L}^{L} \mathrm{d}x\,\hat\Psi^\dagger_\sigma(x) H^{(\lambda)}_\sigma \hat\Psi_\sigma(x) +\\
 + g\int_{-L}^{L}\mathrm{d}x\,
 \hat\Psi^\dagger_\uparrow(x)\hat\Psi^\dagger_\downarrow(x)\hat\Psi_\downarrow(x)\hat\Psi_\uparrow(x),
\end{split}
\end{equation}
where the dimensionless interaction strength is $g=g_{\mathrm{1D}} \sqrt{m_{\downarrow}/(\omega\hbar^3)}$.
All integrations are performed over the whole space where the particles could be present, i.e. in the region between the walls $(-L,L)$. The field operator $\hat\Psi_{\sigma}(x)$ annihilates fermions of spin $\sigma$ at a position $x$. The quantum fields obey canonical anti-commutation relations for same spin particles $\left\{\hat{\Psi}_\sigma(x),\hat{\Psi}^\dagger_\sigma(x')\right\}=\delta(x-x')$ and $\left\{\hat{\Psi}_\sigma(x),\hat{\Psi}_\sigma(x')\right\}=0$. In contrast, due to the fundamental distinguishability of opposite spin fermions explained before, the final result and the values of calculated observables do not depend on the choice of the commutation relations for opposite spin operators \cite{Weinberg}. However, as commonly used for distinguishable particles, we assume commutation of the field operators in this case, $\left[\hat{\Psi}_\uparrow(x),\hat{\Psi}^\dagger_\downarrow(x')\right]=\left[\hat{\Psi}_\uparrow(x),\hat{\Psi}_\downarrow(x')\right]=0$. Note that in the Hamiltonian (\ref{eq:ham}) there are no terms that change the number of particles of a given flavour. As a consequence, the total number of fermions of a given flavour, $\hat N_{\sigma} = \int_{-L}^{L} \hat\Psi_{\sigma}^\dag (x) \hat\Psi_{\sigma}(x) \mathrm{d}x$, commutes with the many-body Hamiltonian (\ref{eq:ham}). This property of the model corresponds to realistic experimental situations where the number of particles can be controlled with an extreme precision\cite{serwane2011deterministic,wenz2013fewToMany}. From the numerical point of view, it enables one to perform a complete analysis of the Hamiltonian independently in each of the subspaces corresponding to a given number of particles.

\section{Exact diagonalization approach} \label{Sec:ExactDiag}
 The ground-state properties of the system are studied straightforwardly within an exact diagonalization approach for the many-body Hamiltonian.  Recently, the method has been successfully used for equal mass fermions confined in a harmonic trap \cite{SowinskiGrass2013FewInteracting,Sowinski2015Pairing,Sowinski2015SlightlyImbalanced} as well as for fermions of different masses \cite{Pecak2016Separation}. First we decompose the field operators $\hat{\Psi}_{\sigma}(x)$ into the basis of the eigenfunctions of the corresponding single-particle Hamiltonians (\ref{eq:spHams}), 
\begin{equation}\label{eq:FieldOperator}
\hat{\Psi}_{\sigma} (x) =  \sum_n \phi^{(\lambda)}_{n\sigma}(x) \hat{a}_{n \sigma},
\end{equation}
where an operator $\hat{a}_{n \sigma}$ annihilates a fermion of the $\sigma$-type in level $n$, i.e. a fermion in a single-particle state described by the wave function $\phi^{(\lambda)}_{n\sigma}(x)$. Note that for simplicity we omit the superscript $\lambda$ in the definition of an annihilation operator since it should not lead to any confusion.

The expansion \eqref{eq:FieldOperator} is exact provided the sum runs over all $n$. In practice, to perform numerical calculations we cut the summation at a value $n_{\mathrm{cutoff}}$ chosen in such a way that the final results do not change significantly when $n_{\mathrm{cutoff}}$ is increased. Of course, for stronger interactions $g$, more single-particle levels should be taken into account to achieve the convergence. For example, for $g=4$, $N_{\downarrow}=2$, and $N_{\uparrow}=3$, we use 12 single-particle eigenstates for each component, i.e. the dimension of the many-body Hilbert space is 14520.

With the expansion \eqref{eq:FieldOperator} the Hamiltonian (\ref{eq:ham}) can be rewritten in the form
\begin{equation}\label{eq:sqHam}
 \hat{\cal H} = \sum_{\sigma} \sum_n E^{(\lambda)}_{n\sigma} \hat{a}_{n\sigma}^{\dag}\hat{a}_{n\sigma} + \sum_{ijkl} U^{(\lambda)}_{ijkl} \hat{a}_{i\uparrow}^{\dag} \hat{a}_{j\downarrow}^{\dag} \hat{a}_{k\downarrow} \hat{a}_{l\uparrow},
\end{equation}
where $E^{(\lambda)}_{n\sigma}$ is a single-particle energy. The interaction energy has the form
\begin{equation}
 U^{(\lambda)}_{ijkl} = g\int\!\!\mathrm{d}x\,\,\bar\phi^{(\lambda)}_{i\uparrow}(x)\bar\phi^{(\lambda)}_{j\downarrow}(x)\phi^{(\lambda)}_{k\downarrow}(x)\phi^{(\lambda)}_{l\uparrow}(x).
\end{equation}  
The Hamiltonian \eqref{eq:sqHam} is represented using all its matrix elements between states belonging to the Fock space of all the possible many-body configurations of $N_\uparrow$ and $N_\downarrow$ particles occupying the first $n_{\mathrm{cutoff}}$ single-particle orbitals. Finally, an exact diagonalization of the matrix obtained is performed using the Implicitly Restarted Arnoldi method \cite{ARPACK1998Sorensen} available in the ARPACK Fortran library. This allows us to find the many-body ground-state of the system $|\mathtt{G}_0\rangle$, several excited states $|\mathtt{G}_i\rangle$ and their eigenenergies ${\cal E}_i$. In this way complete information about the structure of the many-body ground-state (and excited states if necessary) can be obtained. In what follows, we concentrate on the simplest quantity that can be measured experimentally in a straightforward way, namely the single-particle density profile (normalized to the number of particles in a given flavour)
\begin{equation} \label{DensProf}
\rho_\sigma(x) = \langle \mathtt{G}_0|\hat\Psi_\sigma^\dagger(x)\hat\Psi_\sigma(x)|\mathtt{G}_0\rangle.
\end{equation}

\section{Many-body spectral properties} \label{Sec:Spectral}
\begin{figure}
 \resizebox{0.48\textwidth}{!}{%
 \includegraphics{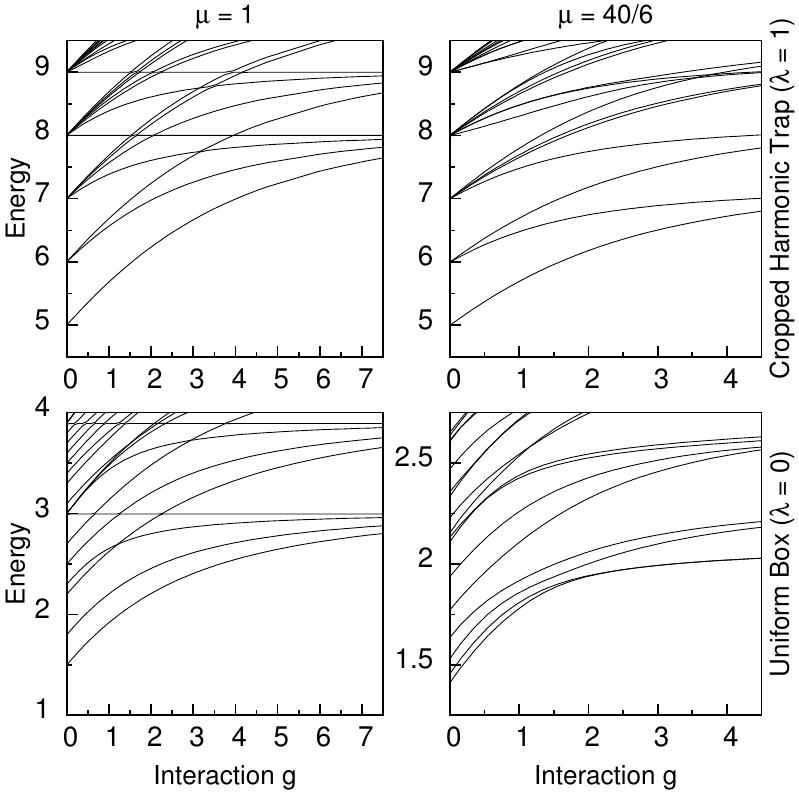}
 }
 \caption{Spectra of the system consisting of $N_{\downarrow}=3$ and $N_{\uparrow}=1$ fermions as a function of the dimensionless interaction strength $g$. The top row corresponds to a harmonic oscillator potential and the bottom row to a box trap potential. Quasi-degenerate energy bands seen in the left column split up when mass imbalance $\mu \neq 1$ is introduced (right column). The energy is given in the natural units of harmonic oscillator, $\hbar\omega$.
 \label{fig:fig2} }
\end{figure}

First let us study how the spectral properties of the many-body Hamiltonian are affected by the shape of the external potential $\lambda$ and mass ratio $\mu$. The results for a harmonic oscillator (shown in the upper panels of Fig. \ref{fig:fig2}) were recently discussed with all details in \cite{Pecak2016Separation}. There, it was shown that along with an increasing mass ratio $\mu$ the quasi-degeneracy of the many-body spectrum is split in the limit of strong interactions. This is caused by the lifting of some global symmetries of the Hamiltonian that are present only in equal mass systems. As a consequence, separation between spin components appears in the ground-state of the system for strong enough interactions, i.e. the heavier particles always concentrate in the middle of the trap and the cloud of light particles is divided into two parts and pushed out from the center. It was noticed that the separation of the density profile induced by a mass imbalance always displays the same features regardless of the number of particles in both flavours. 

The situation is very similar for the case of a uniform box potential ($\lambda=0$). The many-body spectrum becomes more complicated for strong interactions whenever different masses of constituents are introduced (bottom right panel of Fig. \ref{fig:fig2}). The main qualitative difference appears in the limit of vanishing interactions -- in contrast to the case of the harmonic oscillator, the spectrum of the uniform box with noninteracting particles changes with $\mu$. This is a direct consequence of the form of the single-particle energies (\ref{EnergyHO}) and (\ref{EnergyBox}).

At this point it is also worth noting that for an equal mass system, and for any confinement, there exist many-body eigenstates that are absolutely insensitive to the interaction strength (seen as horizontal lines in the left panels of Fig. \ref{fig:fig2}). These states, commonly named after Girardeau \cite{Girardeau1960}, are straightforwardly constructed using a single Slater determinant of $N_\uparrow+N_\downarrow$ single-particle orbitals. Such wave functions are antisymmetric under the exchange of the positions of any two fermions, regardless of their spin. Thus they are the eigenstates of the interaction part of the Hamiltonian. This construction of completely antisymmetric states can only be adopted for equal mass systems since only then the single-particle orbitals are the same for both flavours. This is the reason why the Girardeau states are not present in the right panels of Fig. \ref{fig:fig2}.

\section{Separation of flavours in the uniform box} \label{Sec:Uniform}
\begin{figure*}
\resizebox{0.68\textwidth}{!}{%
 \includegraphics{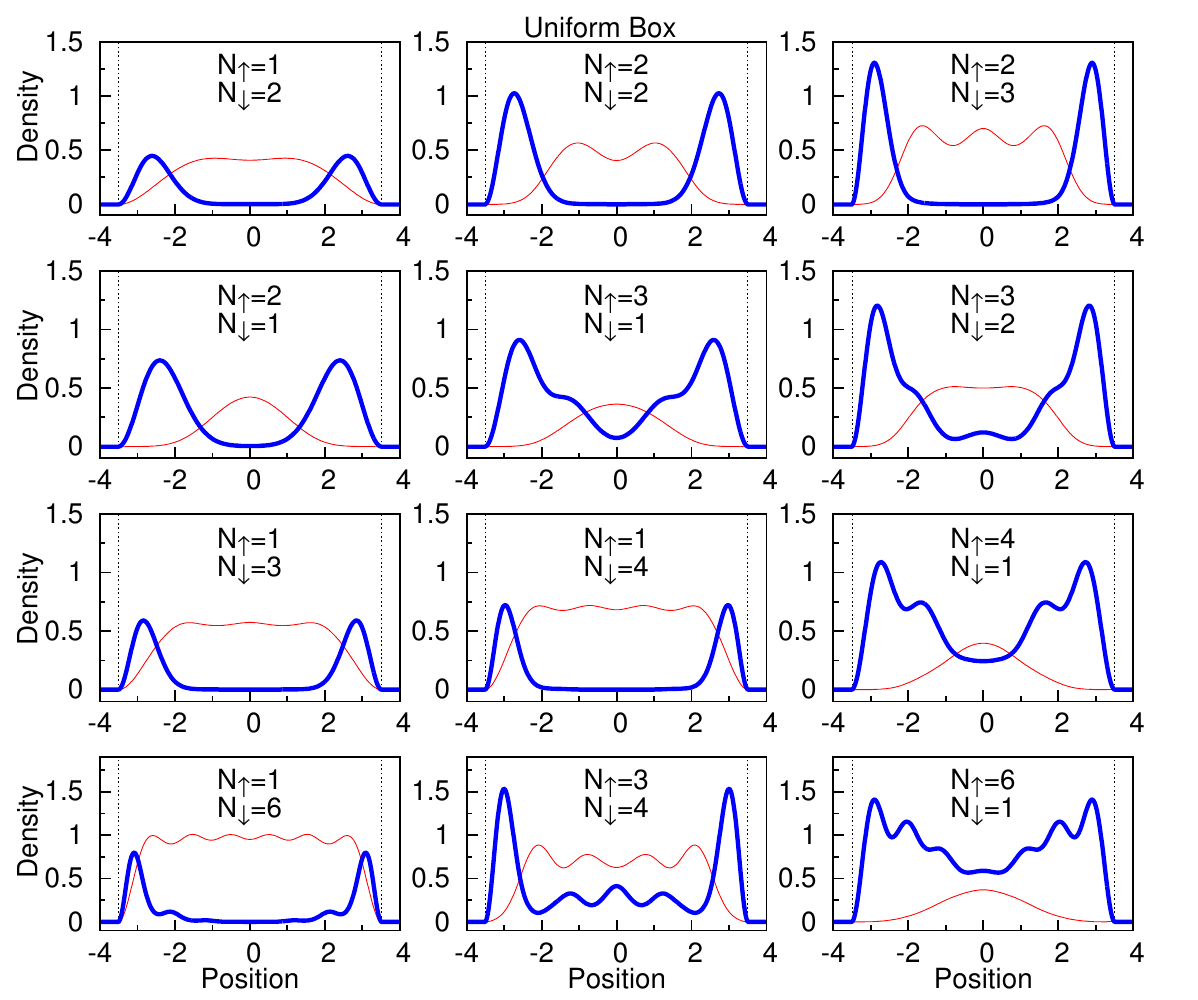}
 }
 \caption{Single-particle densities $\rho_\uparrow(x)$ (thick blue line, heavier flavour) and $\rho_\downarrow(x)$ (thin red line, lighter flavour) calculated in the ground-state of the system for different numbers of fermions with $\mu=40/6$ and strong interaction $g=4$ confined in a box trap. The black vertical lines correspond to the walls of the box trap. In contrast to the separation induced by the mass difference for a harmonic potential \cite{Pecak2016Separation}, in this case the separation always occurs in the heavier fraction, independently of the way the fermions are distributed between flavours. In particular, the separation is present also for an equal number of fermions $N_\uparrow=N_\downarrow$. The positions and the densities are measured in units of $\sqrt{\hbar/(m_{\downarrow} \omega)}$ and $\sqrt{m_{\downarrow} \omega/\hbar}$, respectively.
 \label{fig:fig3}}
\end{figure*}

\begin{figure}
 \resizebox{0.48\textwidth}{!}{%
 \includegraphics[scale=1.2]{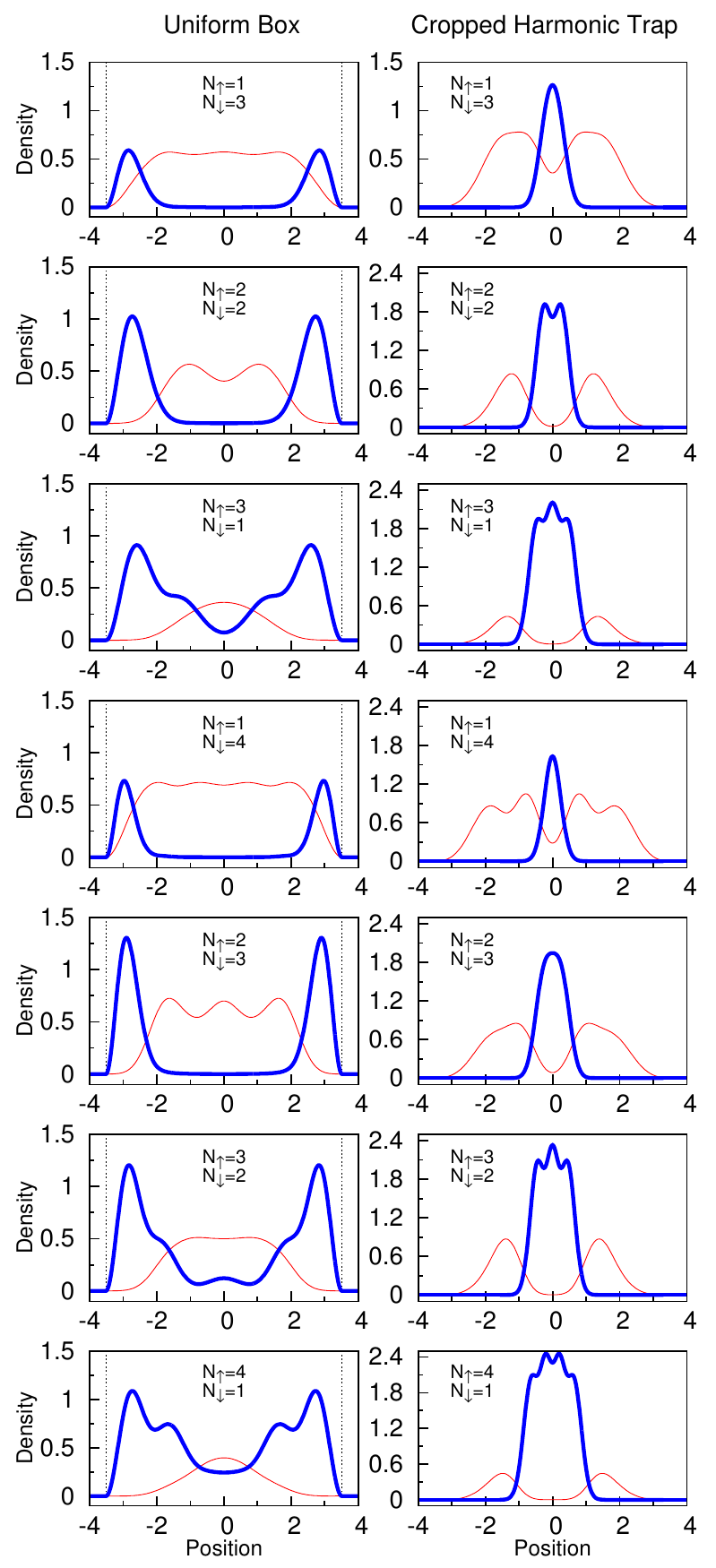}
 }
 \caption{Comparison of different separation scenarios for fermions with different mass ($\mu=40/6$) driven by different shapes of the confinement, in the limit of strong interaction $g=4$: in the uniform box (left panels) and cropped harmonic oscillator (right panels). The thick blue and thin red lines represent the single-particle density profiles for heavy and light components, respectively. Note that, independently on the number of particles in a given flavour, the separation is always present in the heavier (for the uniform box) or the lighter (for the harmonic oscillator) component. The positions and the densities are measured in units of $\sqrt{\hbar/(m_{\downarrow} \omega)}$ and $\sqrt{m_{\downarrow} \omega/\hbar}$, respectively. \label{fig:fig4} }
\end{figure}
As mentioned previously, in harmonic confinement, the mass difference between fermions of different flavours leads to the separation of density profiles of opposite species for strong enough repulsions. In the case of the uniform box potential a separation of the density also occurs in the system. However, in this case, the separation is present always in the heavier component (see Fig. \ref{fig:fig3}).
The direct reason why a mass difference acts differently for different confinements can be explained intuitively via energetic arguments. As mentioned previously, in the uniform box case, the single-particle wave functions are exactly the same for both components and they are completely independent of mass difference. Therefore, the part of the energy cost for exciting a particle to a higher state that comes from the interaction, is independent of the flavour. The only difference in energies comes from the single-particle part of the Hamiltonian. The energy cost for exciting heavier particles is smaller (see eq. (\ref{EnergyBox})) and therefore the separation in heavier component is favoured. This argumentation is completely opposite to that in the case of harmonic confinement (see \cite{Pecak2016Separation} for details) and therefore the separation is governed by an opposite rule.

These intuitive pictures and arguments are confirmed by our numerical calculations. In Fig.~\ref{fig:fig3}, the single-particle density is plotted for a strongly interacting system of two species characterized by a mass ratio of $\mu=40/6$. We have checked that the separation occurs in the strong interaction limit for any number of particles up to seven. From the same calculations we have seen that for mass ratios $\mu$ closer to $1$, a much stronger interaction is needed to create the separation in density profiles. This observation is also in accordance with our intuitive picture, i.e. for almost equal masses neither component is favored and much stronger interactions are needed to break the symmetry and support separation. 

For completeness, in Fig. \ref{fig:fig4}, we show a comparison of separations for the two confinements considered, i.e. the uniform box (left panels) and the harmonic trap (right panels). Matching plots are obtained for the same number of particles in both components and the same interaction strengths. From this comparison it is obvious that the separation mechanism induced by a mass imbalance acts completely differently in the two cases.

\section{Comparison to the equal mass system} \label{Sec:EqualMass}
\begin{figure}
 \resizebox{0.48\textwidth}{!}{%
 \includegraphics[scale=1.2]{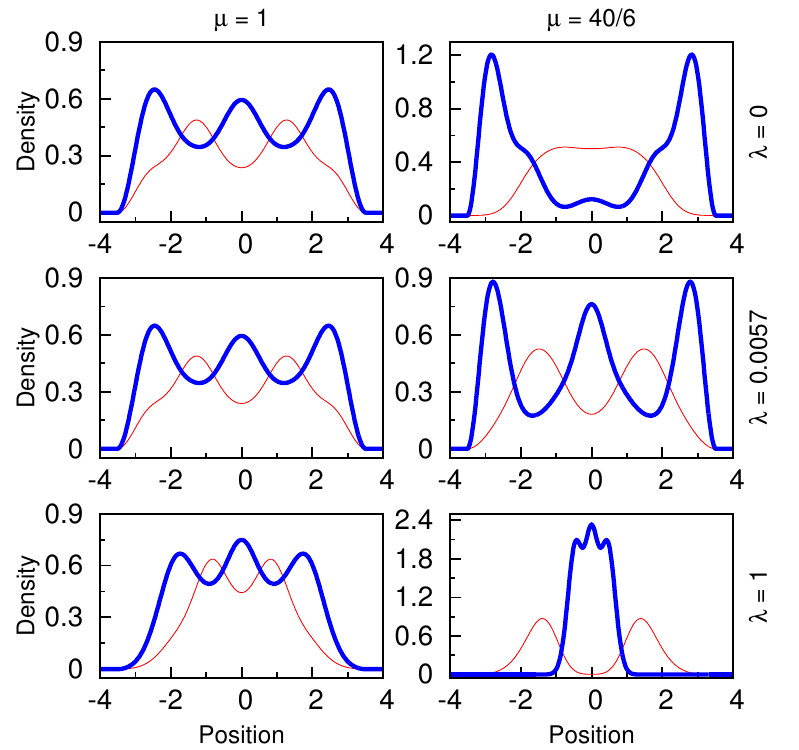}
 }
 \caption{Single-particle densities $\rho_\uparrow(x)$ (thick blue line, heavier flavour) and $\rho_\downarrow(x)$ (thin red line, lighter flavour) calculated in the ground-state of the system of $N_{\uparrow}=3$ and $N_{\downarrow}=2$ in the limit of strong interaction, $g=4$. The positions and the densities are measured in units of $\sqrt{\hbar/(m_{\downarrow} \omega)}$ and $\sqrt{m_{\downarrow} \omega/\hbar}$, respectively.
 \label{fig:fig5} }
\end{figure}

Before we analyze the transition in the ground-state between the two orderings described above, let us compare the situation to the case when both flavours have the same mass. It is known that in this case the separation can be induced only by a difference in the number of particles, $N_\uparrow-N_\downarrow$. This arises directly from the general symmetry under global exchange of both families of particles. As consequence, whenever $N_\uparrow=N_\downarrow$, both flavours have the same single-particle density profile and no separation of the density profile can be observed. The situation is modified when the system is slightly imbalanced in the number of particles. As an example we concentrate on the system with $N_\uparrow=3$ and $N_\downarrow=2$ particles. As seen in the left panels of Fig. \ref{fig:fig5}, characteristic alternating oscillations in the densities of the ground-state are built in the limit of very strong repulsions and both components take on an antiferromagnetic ordering. It is seen that an alternating ordering is present in the system independently of the shape of the external potential. This generalizes the result obtained recently for harmonic confinement for finite interactions\cite{Lindgren2014Fermionization}, and extends the results obtained for infinite interactions \cite{Volosniev2014Strongly1D,ReimannSantosChain2014}.

The situation is very different whenever different masses of the components are introduced (see the right panels in Fig. \ref{fig:fig5}). Under harmonic confinement, the heavier particles concentrate in the middle and the lighter ones are pushed out from the center. For the case of the uniform box, heavier particles are located in the vicinity of the walls and lighter ones are in the middle. The middle plot shows a generic situation for an intermediate confinement shape. It suggests that in this case, separation is not present. One should remember however, that the plot is obtained for strong but finite interactions. Our numerical calculations, performed for many different arrangements (different confinements and different numbers of particles) show that for an arbitrary confinement in the range $0\leq \lambda\leq 1$ there exists some critical interaction strength above which one of the two separation types occurs in the system. One can imagine that for infinite interactions any few-fermion system with imbalanced mass reveals spatial separation in single-particle distributions. The only question is if the separation is built in the heavier or the lighter component. The answer is directly related to the shape of the confinement.  From the above analysis it follows that the system undergoes some kind of transition between different separations in the limit of infinite interactions, which is driven by an adiabatic change of the potential. As explained below, the properties of this transition can be understood with methods well known from the theory of quantum phase transitions. 

\section{The transition driven by the shape of the trap} \label{Sec:Transition}
\begin{figure}
 \resizebox{0.48\textwidth}{!}{%
 \includegraphics[scale=1.2]{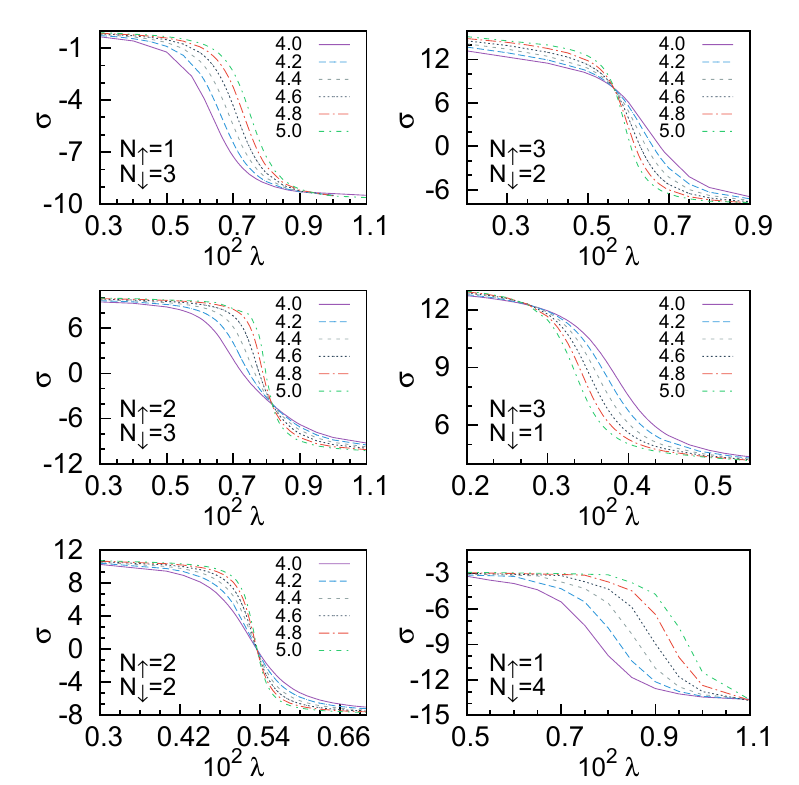}
 }
 \caption{The second moment $\sigma$ of the magnetization distribution (\ref{Sigma}) as a function of the shape of the confinement for different interaction strengths (from $g=4$ to $g=5$). Each plot corresponds to given numbers of particles in both flavours. Note that in extreme confinements, $\sigma$ saturates to a well defined value, while it changes rapidly in the vicinity of the transition point. The second moment $\sigma$ is given in the natural units of a harmonic oscillator, $\hbar/(m_{\downarrow} \omega)$. \label{SecondMoment} }
\end{figure}

As explained above, for the two extreme cases of a uniform box and a harmonic trap, the density separation induced by the mass imbalance is of an opposite kind. Depending on the spectrum of the single-particle Hamiltonians, heavier or lighter particles are pushed out from the center for sufficiently large repulsions between particles. In the framework of our model it is possible to study the transition between these two orderings induced by an adiabatic change of the shape. To make this analysis not only qualitative but also quantitative one should introduce some quantity which indicates the kind of ordering. The choice is obviously not unique, however it is quite natural to concentrate on a magnetization-like distribution defined as follows: 
\begin{equation} \label{Magnet}
 M(x) = \rho_{\uparrow}(x) - \rho_{\downarrow}(x).
\end{equation}
It is quite natural that this distribution has an opposite behavior whenever heavier or lighter particles are pushed out from the center of the trap. Since the distribution is normalized to the difference of the total number of particles, $N_\uparrow-N_\downarrow$, and also because it is symmetric under spatial reflections with respect to $x=0$, therefore the first distinction between the two orderings being considered is manifested by the value of the second moment of the distribution 
\begin{equation} \label{Sigma}
 \sigma = \int_{-L}^{L}\!\mathrm{d}x\,x^2\,M(x).
\end{equation}
In Fig. \ref{SecondMoment} we show the dependence of $\sigma$ on the trap parameter $\lambda$ for different numbers of particles and different interactions. It is seen in the two extreme confinements, that $\sigma$ saturates to the two completely distinct values corresponding to the two different orderings. It means that $\sigma$ plays the role of an order parameter and can be used as an indicator of a given ordering. As long a given ordering is present in the system, the parameter $\sigma$ is almost constant. Near the transition point, however, (a point that is different for different numbers of particles) its value rapidly changes. Moreover, for stronger interactions, the transition is more sharp. Therefore, one can anticipate that for infinitely strong repulsions a characteristic 'step-like' function is obtained. All the above points mean that the transition between orderings appearing for strong interactions has many of the properties of a phase transition \cite{Sowinski2015Criticality,Gu2008Fidelity} and it can be analyzed with the methods known from the theory of quantum phase transitions \cite{Sachdev2011QPT,NewmanBarkema1999MC}. Here, the roles of the order parameter and the parameter of control are played by the second moment of the magnetization-like distribution $\sigma$, and the shape of the trap $\lambda$, respectively. From this point of view, the thermodynamic limit is mimicked by the limit of infinitely strong repulsions between particles.

To characterize the transition between different orderings one should study not only the behavior of the order parameter but also its derivatives. Naturally, the most important of these is the lowest derivative that is divergent at the transition point. Our numerical results suggest that, in the case studied, the first derivative of $\sigma$ has this property in the limit of infinite interactions. In analogy to the physics of phase transitions this quantity has all the properties of the susceptibility since it measures changes of magnetization under small changes of the parameter of control
\begin{figure*}
 \resizebox{0.68\textwidth}{!}{%
 \includegraphics[scale=1.2]{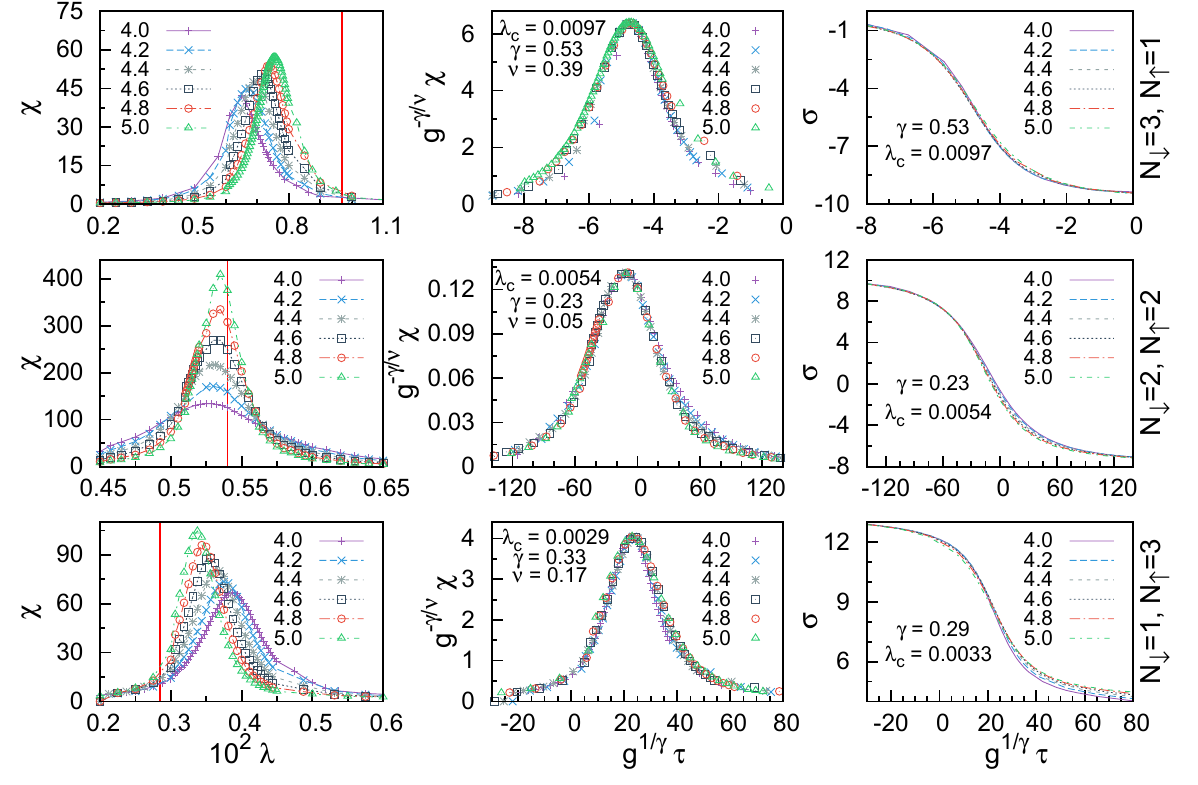}
 }
 \caption{Scaling properties of a few-body system. Left panels: Susceptibility $\chi$ as a function of the shape of the confinement $\lambda$ for different values of interactions and different number of particles. A characteristic peak  of the susceptibility, whose height increases with $g$, is clearly visible. The vertical red line corresponds to the critical value $\lambda_c$ obtained after extrapolation of the results to infinite repulsion. Middle panels: Rescaled susceptibility as a function of a rescaled confinement shape parameter obtained after adopting the data-collapse method. Note that all data points collapse to one well defined curve. Right panels:  The second moment of the distribution $\sigma$ when the same scaling procedure is performed for the trap shape parameter. The susceptibility $\chi$ is given in the natural units of a harmonic oscillator, $\sqrt{\hbar/(m_{\downarrow} \omega)}$. \label{Susc} }
\end{figure*}
\begin{equation}
\chi(\lambda) = \frac{\mathrm{d} \sigma(\lambda)}{\mathrm{d} \lambda}.
\end{equation}
We numerically calculate the susceptibility $\chi$ for different numbers of particles and for different interactions $g$ (examples for $N_\downarrow+N_\uparrow=4$ are shown in the left panels of Fig. \ref{Susc}). The susceptibility calculated in this way has a natural behavior well known from the theory of phase transitions. Its maximum grows with interactions along with a small shift of its position. One can anticipate that for infinitely large interactions the susceptibility will be divergent at the position of the transition point. This behavior is a direct consequence of the sharpening of the $\sigma$ function. The analogy with the theory of quantum phase transitions is seen to be even closer when we adopt the well known finite-size scaling method to determine the position of the transition point in the limit of infinite interactions. First we assume that the order parameter defined by $\sigma$ has some natural scaling in the vicinity of the transition point $\lambda_c$, i.e. it is a homogeneous function of its relevant parameters: interaction strength $g$ and the normalized shape of the trap defined as $\tau=(\lambda-\lambda_c)/\lambda_c$. Consequently, the same property is shared by all its derivatives. Regarding the susceptibility, this means that there exists one universal function $\tilde\chi(\xi)$ that determines the shapes of all susceptibilities for different confinements and interaction strengths. To make the analogy to the theory of quantum phase transitions as close as possible we assume the following scaling ansatz \cite{Gu2008Fidelity,Sowinski2015Criticality}:
\begin{equation}
 \chi (\tau,g) = g^{\gamma / \nu} \tilde{\chi} (g^{1/\nu} \tau),
\end{equation}
where $\nu$ and $\gamma$ are appropriate critical exponents of the model. If the assumption of the scaling property of the susceptibility is correct, then there exists an appropriate choice of critical exponents for which all numerical data points form the one universal curve determined by $\tilde{\chi}$. To show that indeed our system has this scaling property we performed appropriate numerical calculations based on the {\it data-collapse method} (for details see for example \cite{Sowinski2015Criticality,Gu2008Fidelity,NewmanBarkema1999MC}). As the result of this numerical approach, we obtain the plots shown in the middle panels of Fig. \ref{Susc}.

It is clearly visible that after appropriate scaling, all the curves for a given system collapse to one universal curve for a large range of normalized potential shapes $\tau$. The position of the transition point $\lambda_c$, as well as critical exponents, are presented in the legend of their corresponding plots. Note that, depending on the number of particles, different values of the critical parameters are obtained. Finally, to make the presentation complete, in the right panels of Fig. \ref{Susc} we show the second moment of the distribution $\sigma$ when the same scaling transformation is performed. It is seen that also in this case all data points collapse to one universal curve. Together, all these results suggest that the transition between different orderings driven by an adiabatic change of the shape of the trap, in the limit of very strong interactions, has many properties similar to those known from the theory of quantum phase transitions. This means that in the limit of infinite interactions, for a given shape of the trap, the system has a well established ordering. In the vicinity of the transition point, the system undergoes a rapid transition -- single-particle densities change to form a new ordering.  

\section{Conclusions} \label{Sec:Conclude}
To conclude, in this article we have discussed the properties of several fermions confined in a one-dimensional trap in the very strong interaction limit. We show that the mass difference between components, independently of the confinement's shape, always leads to a spatial separation between flavours. However, the nature of the separation depends on the shape, i.e. for a given shape the density profile of either lighter or heavier particles is split into two parts and pushed out from the center of the trap. This observation subsequently led us to the concept of a transition between orderings driven by the shape of the trap. We show that this transition has many properties in common with standard quantum phase transitions, and can be similarly analyzed within the finite-size scaling framework. In this way we find critical shape values for different numbers of particles for which the system undergoes transitions and we have estimated the relevant critical exponents for these transitions. It is worth noticing that in the case of one-dimensional systems, typically smooth crossovers rather than rapid transitions between different phases are suspected. From this point of view the transition predicted here is quite a rare and interesting phenomenon.

\acknowledgments
The authors thank P Deuar for careful reading of the manuscript and helpful language suggestions. This work was supported by the (Polish) Ministry of Science and Higher Education, Iuventus Plus 2015-2017 Grant No. 0440/IP3/2015/73.

\end{document}